\documentclass[%
 reprint,
 superscriptaddress,
 footinbib,
 amsmath,amssymb,
 aps,
 prl,
]{revtex4-1}

\usepackage{graphicx}
\usepackage{dcolumn}
\usepackage{bm}

\newcommand{\eref}[1]{Eq.~(\ref{#1})}

\begin{document}

\preprint{APS/123-QED}

\title{
A Simple Explanation for the Observed Power Law Distribution of Line Intensity\\
in Complex Many-Electron Atoms
} 

\author{Keisuke Fujii}
 \email{fujii@me.kyoto-u.ac.jp}
 \affiliation{%
Department of Mechanical Engineering and Science,
Graduate School of Engineering, Kyoto University,
Kyoto 615-8540, Japan}
\affiliation{%
Max-Planck-Institut f\"ur Kernphysik, Saupfercheckweg 1, 69117 Heidelberg, Germany
}

\author{Julian C. Berengut}%
\affiliation{%
School of Physics, University of New South Wales, New South Wales 2052, Australia}
\affiliation{%
Max-Planck-Institut f\"ur Kernphysik, Saupfercheckweg 1, 69117 Heidelberg, Germany
}%

\date{\today}

\begin{abstract}
It has long been observed that the number of weak lines from many-electron atoms follows a power law distribution of intensity.
While computer simulations have reproduced this dependence, its origin has not yet been clarified.
Here we report that the combination of two statistical models --- an exponential increase in the level density of many-electron atoms and local thermal equilibrium of the excited state population --- produces a surprisingly simple analytical explanation for this power law dependence.
We find that the exponent of the power law is proportional to the electron temperature. This dependence may provide a useful diagnostic tool to extract the temperature of plasmas of complex atoms without the need to assign lines.
\end{abstract}

\maketitle



It has long been known that the number of weak lines emitted by many-electron atoms in plasmas follows an intensity power law.
In 1982 Learner pointed out this law for the first time when
measuring emission lines from a hollow cathode lamp containing iron atoms~\cite{Learner1982}. He observed that the number density of lines with a given intensity $I$, $\rho_I(I)$,
exhibits a power law dependence on $I$~\footnote{Note that although he used a different base, 2 for the intensity and 10 for the line number, we use $e$ as the base and present the converted value by $b \log 10 / \log 2 - 1$ for later convenience, where $b$ is the original value, -0.15.},
\begin{equation}
    \label{eq:learner}
    \rho_I(I) \propto I^{-1.50}.
\end{equation}
He also reported that $\rho_I(I)$ in different wavelength regions all follow this power law  with the same exponent, indicating an ergodic property of the emission line distribution \cite{Learner1982}.

This work has stimulated much discussion.
A theoretical study by Scheeline showed that this power law does not hold for hydrogen atom spectra \cite{Scheeline1986}.
In contrast, the emission spectrum from arsenic, which has a much more complex electronic structure than hydrogen, shows an intensity distribution closer to the power law, but with a different value of the exponent \cite{Scheeline1971}.
Bauche-Arnoult and Bauche reported a simulation with a collisional-radiative model for neutral iron atom and demonstrated that the power law dependence is again reproduced \cite{Bauche-Arnoult1997}.
Their exponent was 17--25 \% smaller than the Learner's value but the reason was not clarified.

Pain recently reviewed this power law dependence problem and presented a discussion regarding fractal dimension and quantum chaos \cite{Pain2013}.
According to his discussion, the line strength distribution evaluated under the fully quantum-chaos assumption does not explain Learner's law.
As presented in his review \cite{Pain2013}, as well as in the book \cite{Bauche2015}, the origin of this power law is still not understood, despite almost 40 years passing since the first report.

In this Letter, we present a surprisingly simple explanation of Learner's law. We assume local thermal equilibrium of the excited state population, and an exponential increase in the level density of complex atoms, which has been reported in several many-electron atoms and ions (e.g.~\cite{Flambaum1998, Dzuba2010}). Combining these, we show below that the number of levels with a given population follows a power law distribution.
An assumption of independently and identically distributed radiative transition rates then directly gives Learner's law in the form
\begin{equation}
    \label{eq:final}
    \rho_I(I) \propto I^{-2kT_\mathrm{e} / \epsilon_0 - 1},
\end{equation}
where $k$ is Boltzmann's constant and $T_\mathrm{e}$ is electron temperature in the plasma.
$\epsilon_0$ is a scale parameter representing the level density growth rate against the excited energy (see \eref{eq:exponential_law} for its definition), which can be estimated either from the experimentally derived energy levels or from \textit{ab initio} atomic structure calculations \cite{Dzuba2010}.

Plasma spectroscopy has been developed from simpler systems, e.g. hydrogen and rare gas atoms.
It is known that comparison between intensity ratios of certain emission lines and collisional-radiative models provide us with information about plasma parameters, such as electron temperature and density \cite{Griem1986, Fujimoto, Sawada1993, Goto2003}.
This requires correct line identifications and accurate atomic data such as energy levels, oscillator strengths, and collision cross sections. 
However, accurate atomic data for open-shell atoms is difficult to obtain despite numerous demands for plasma diagnostics with complex atoms, ranging from laser produced plasmas for extreme-ultraviolet light sources
\cite{Bowering2004, Masnavi2007, Suzuki2012},
heavy-metal-contaminated fusion plasmas \cite{Putterich2008, Murakami2015},
to the emissions found after the $r$--process supernova (kilonova) \cite{Pian2017, Tanaka2018}.
Our result \eref{eq:final} suggests an advantage of using intensity statistics for diagnosing plasmas with many-electron atoms, where accurate \textit{ab initio} simulations of such complex spectra are still difficult with currently available theory and computers.

\begin{figure}[ht]
    \centering
    \includegraphics[width=7.5cm]{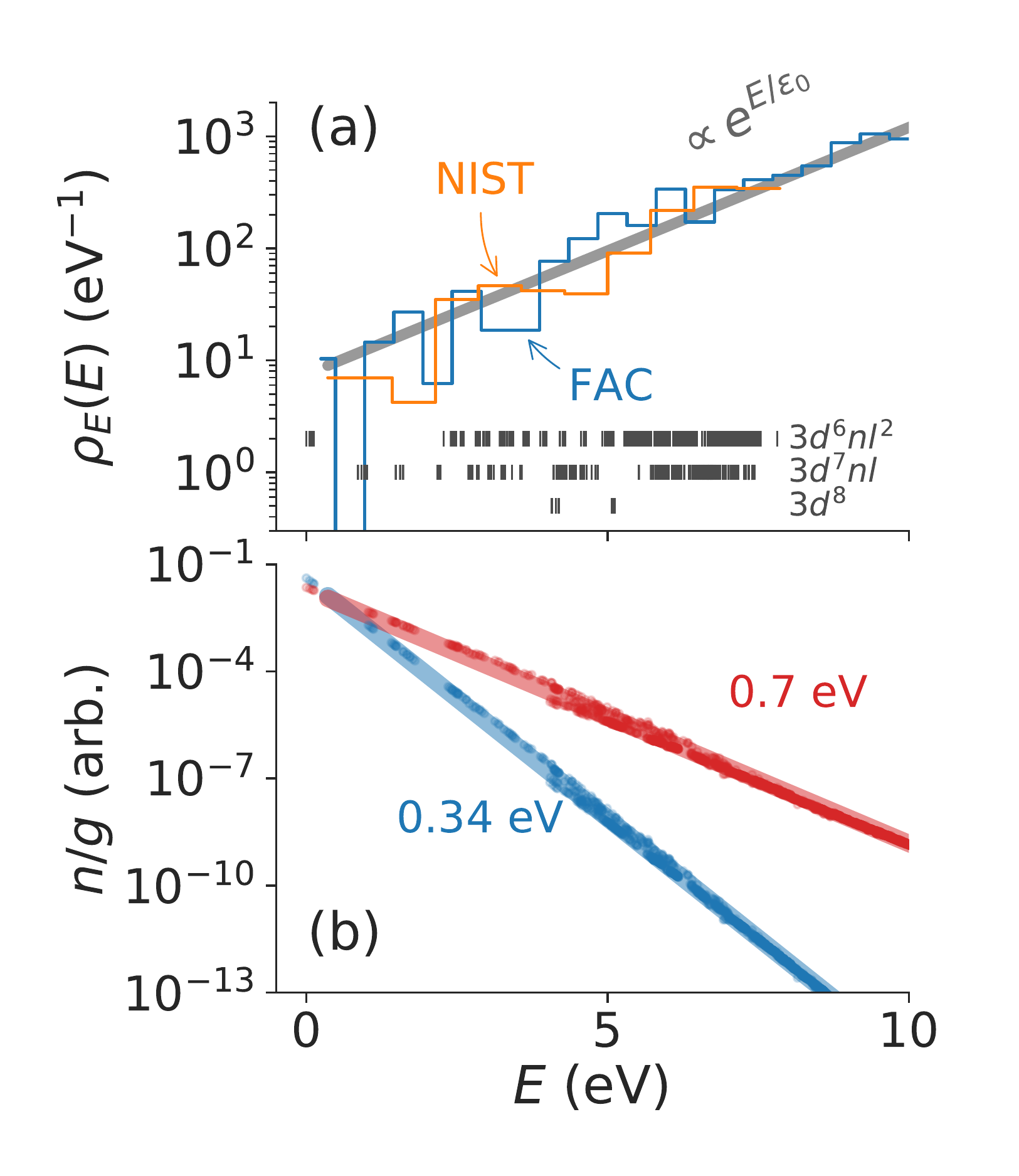}
    \caption{(a) State density of neutral iron against excitation energy.
    The orange histogram is computed from the measurement data compiled in NIST ASD\cite{NIST_ASD}.
    The measured energy levels (846 entries) are shown by the vertical bars in the figure.
    The blue histogram shows the state density computed from FAC \cite{Gu2008}.
    The gray line is an exponential dependence \eref{eq:exponential_law} with $\epsilon_0$ = 1.97 eV, which fits both densities well.
    (b) Population distribution computed using collisional-radiative modelling with $T_\mathrm{e} = $ 0.34~eV (blue points) and 0.7~eV (red points) and electron density $n_\mathrm{e} = 10^{20} \mathrm{m^{-3}}$.
    The blue and red bold lines present the Boltzmann distribution (\eref{eq:boltzmann}) with effective temperatures 0.32~eV and 0.61~eV, respectively.
    }
    \label{fig:level_energy}
\end{figure}

Next we derive \eref{eq:final}, illustrating our assumptions using Learner's example of neutral iron.
Figure \ref{fig:level_energy}(a) shows the level density of neutral iron, $\rho_E(E)$, the number of levels with given excited energy $E$.
This state density is evaluated from the measured energy levels taken from Atomic Spectral Database by National Institute of Standards and Technology \cite{NIST_ASD}.
The observed energy levels are shown by the vertical bars in the figure.

It is well known that the excited level density in the quantum many-body system increases nearly exponentially.
One simple but common approximation is \cite{TerHaar1949, VonEgidy1986, Dzuba2010}
\begin{equation}
    \label{eq:exponential_law}
    \rho_E(E) \propto \exp\left(\frac{E}{\epsilon_0}\right),
\end{equation}
where $\epsilon_0$ is an atom-specific energy scale, which depends on the number of active electrons and the number, degeneracy, and distribution of single particle states in the atom \cite{Dzuba2010}. It can be calculated numerically, derived from experimental energy levels, or estimated using combinatorics.
Dzuba \emph{et al.} presented that for open-$d$- or $f$-shell atoms the state density follows \eref{eq:exponential_law}, at least below the ionization energy \cite{Dzuba2010}.
For neutral iron, we find \eref{eq:exponential_law} well represents the level density with $\epsilon_0 \approx 1.97 \pm 0.04$ eV as indicated by the solid line in Fig.~\ref{fig:level_energy}(a).
Note that this value is obtained by the maximum-likelihood estimation of the simulated energy levels.

Let us assume local thermal equilibrium for the excited state population. The population in state $i$ with energy $E_i$ is given as
\begin{equation}
    \label{eq:boltzmann}
    n_i \propto g_i\exp\left(-\frac{E_i}{kT_\mathrm{e}}\right),
\end{equation}
where $g_i$ is the statistical weight of the state $i$ ($g_i = 2J_i + 1$, where $J_i$ is the total angular momentum quantum number of state $i$).
This equilibrium is valid in plasmas with high electron density and low electron temperature \cite{Fujimoto}.
By substituting \eref{eq:exponential_law} into \eref{eq:boltzmann}, the number of states having the population $n \sim n + \mathrm{d}n$ can be written as,
\begin{align}
    \rho_n(n)\mathrm{d}n
    & = \rho_E(E) \mathrm{d}E \propto \frac{1}{n}\rho_E(E) \mathrm{d}n \label{eq:derivation1}\\
    & \propto \frac{1}{n} \exp\left(-\frac{kT_\mathrm{e}}{\epsilon_0} \log n\right) \mathrm{d}n \label{eq:derivation2}\\
    & \propto n^{- k T_\mathrm{e} / \epsilon_0 -1} \mathrm{d}n \label{eq:derivation3},
\end{align}
where $\mathrm{d} E$ is the energy interval corresponding to $\mathrm{d}n$, the relation of which can be obtained from \eref{eq:boltzmann}.
We assume in \eref{eq:derivation2} that the statistical weight is distributed uniformly over the energy and therefore we omit it from the equation.
This power law originates from the combination of one exponentially increasing variable and another exponentially decreasing variable.
This is a typical mathematical structure responsible for the emergence of power laws \cite{Simkin2011}.

The emission intensity $I_{ij}$ corresponding to the transition $i \to j$, where $j$ is the lower state, is proportional to the upper state population $n_i$, the transition energy cubed $\omega_{ij}^3 = (E_i - E_j)^3$, and the line strength $S_{ij}$ between $i$ and $j$ states.
In many-electron atoms with sufficient basis-state mixing, i.e., in quantum-chaotic systems, the probability distribution of $S_{ij}$ can be well approximated as uniform and independent, and modeled using the Porter-Thomas distribution $p(S) \propto \frac{1}{\sqrt{2 \pi S_0 S}}\exp(-\frac{S}{2 S_0})$, with a constant $S_0$ \cite{Porter1956, Grimes1983, Bisson1991,flambaum94pra, Flambaum1998}.
This approximation is obtained by modeling the Hamiltonian with a Gaussian orthogonal ensemble.
As this distribution decays considerably faster than the power law in the large $S$ limit, we can safely approximate that $S_{ij}$ is a constant for all pairs of levels.
Therefore, the intensity $I_{ij}$ is approximated as
\begin{equation}
    \label{eq:intensity}
    I_{ij} \propto \omega_{ij}^3 S_0 n_i.
\end{equation}
A more detailed and precise discussion can be found in the Supplementary Material
\footnote{See Supplementary Material for more detailed derivation of \eref{eq:final}, as well as the validity condition of the local-thermal-equilibrium assumption, which includes
Refs. \cite{
Porter1956, Grimes1983, Bisson1991, Flambaum1998,
Mewe1972,
Griem,
McWhirter,
Suzuki2017,Osullivan2015, 
Kawashima2009, Kukushkin2011}}.

The number of emission lines from state $i$ observed in photon energy range $\omega \sim \omega + \mathrm{d} \omega$ is proportional to the number of levels in this energy range, $\rho_E(E_i - \omega) \mathrm{d}\omega$.
By considering the number of emission lines with a given intensity range $I \sim I + \mathrm{d}I$, we arrive at \eref{eq:final},
\begin{align}
\rho_I(I)\mathrm{d}I &= \int_\Omega \rho_E(E - \omega)\rho_E(E) \mathrm{d}\omega \mathrm{d}E
\propto I^{-2kT_\mathrm{e}/\epsilon_0 - 1} \mathrm{d}I
\nonumber
\end{align}
where the integration along $\omega$ is taken over the observed photon energy range, $\Omega$.
Here the variable $E$ is changed to $I$ based on Eqs.~(\ref{eq:boltzmann}) and (\ref{eq:intensity}). The factor 2 newly appears in the exponent of $I$ compared with \eref{eq:derivation3}.

The exponent in \eref{eq:final} does not depend on $\Omega$.
This is consistent with Learner's observation that the emission line density in different wavelength regions all show the power law dependence with the same exponent \cite{Learner1982}.
Learner suggested a relation between the exponent and a constant, $\log_{10}\sqrt{2}$ \cite{Learner1982}.
In contrast, our work clearly indicates a relation with $T_\mathrm{e}$ and the atom-specific constant $\epsilon_0$.

By comparing the exponents in Eqs.~(\ref{eq:learner})~and~(\ref{eq:final}), $T_\mathrm{e}$ in Learner's experiment is estimated as $(1.50 - 1)\epsilon_0/2 \approx 0.49$~eV.
Although in Ref. \cite{Bauche-Arnoult1997} it is claimed that $T_\mathrm{e}$ higher than 0.34 eV is not realistic, $T_\mathrm{e}$ in hollow cathode discharges reported in literature varies from 0.2 eV to 3 eV depending on the cathode element, filler gas pressure, and discharge current \cite{Cottereau1979, Meng2010}.
Therefore, 0.49 eV may not be a surprising value for $T_\mathrm{e}$ in a hollow cathode discharge.

Bauche-Arnoult and Bauche have used $T_\mathrm{e} = 0.34$ eV for their simulation \cite{Bauche-Arnoult1997}, which is smaller than 0.49 eV.
They obtained $-1.392 \pm 0.017$ for the exponent \footnote{This value is also converted from the original value $b = -0.118 \pm 0.005$}, which is consistently smaller in magnitude than Learner's value.
Our above discussion further provides an explanation for one argument in their paper, i.e., the higher the electron temperature, the larger the magnitude of the exponent \cite{Bauche-Arnoult1997}.

\begin{figure}[ht]
    \centering
    \includegraphics[width=7.5cm]{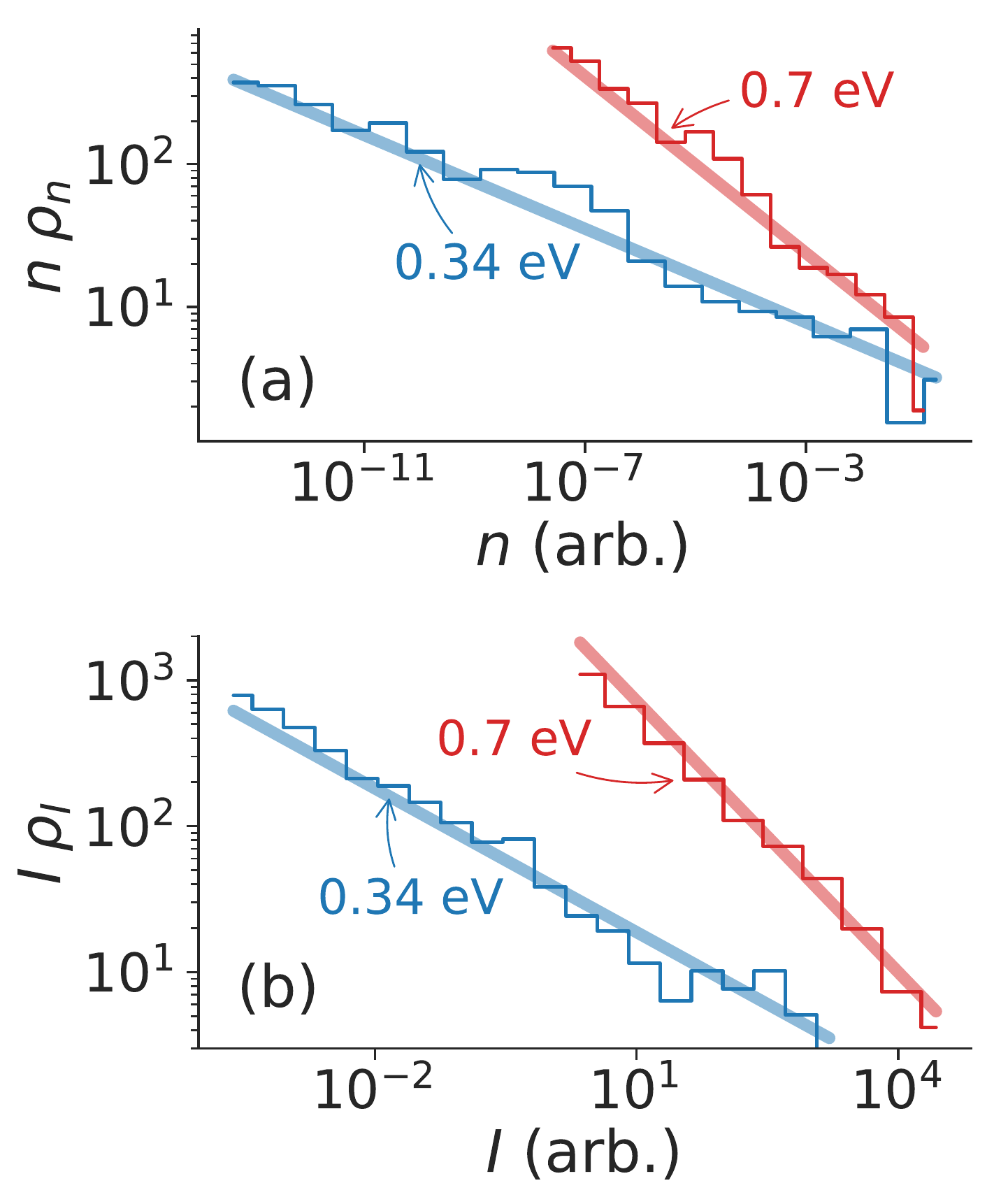}
    \caption{(a) State density distribution $\rho_n(n)$, multiplied by $n$ to aid visualization.
    The blue histogram presents the computed result with $T_\mathrm{e} = 0.34$ eV, while red presents the result with $T_\mathrm{e} = 0.7$ eV.
    Both distributions follow the power law.
    (b) Density distribution of emission lines $\rho_I(I)$, computed by FAC.
    Again, the vertical values are multiplied by $I$ to aid visualization.
    The bold lines in (a) and (b) are not fit results but theoretical models (Eqs.~(\ref{eq:derivation3}) and (\ref{eq:final}), respectively) with the same electron temperatures used in Fig. \ref{fig:level_energy}(b).
    }
    \label{fig:intensity}
\end{figure}

We carry out an \textit{ab initio} simulation of the emission spectrum of neutral iron with the flexible atomic code (FAC) \cite{Gu2008}.
FAC uses the relativistic Hartree-Fock method to calculate the electronic orbitals and
configuration interaction to approximate the electron-electron interaction.
The excited state population and the emission line intensity are evaluated by the collisional-radiative model implemented in FAC, where the steady state of population in the plasma is assumed.
For the collisional-radiative calculation we consider spontaneous emission, electron-impact excitation, deexcitation, and ionization, as well as auto-ionization of levels above the ionization threshold, as elementary processes in plasmas.
These rates are also calculated by FAC.

We assume $T_\mathrm{e} = 0.34$ eV and the electron density $n_\mathrm{e} = 10^{20} \;\mathrm{m^{-3}}$, similar to Bauche-Arnoult and Bauche \cite{Bauche-Arnoult1997}.
We also perform the simulation with $T_\mathrm{e} = 0.70$ eV to observe the $T_\mathrm{e}$-dependence of the exponent.
Note that in the FAC computations we do not explicitly adopt either of our two assumptions, namely the exponential increase of the state density and the local thermal equilibrium of the population.

The state density of a neutral iron atom computed by FAC is shown in Fig. \ref{fig:level_energy}(a) by a blue histogram.
It shows a similar exponential dependence to the measured data, NIST ASD.
Figure \ref{fig:level_energy}(b) shows the excited state population computed by FAC.
Although we do not assume local thermal equilibrium, the population follows the exponential function.
The exponents for $T_\mathrm{e} =$ 0.34 eV and 0.70 eV cases are estimated by the least-squares method to be 0.32 eV and 0.61 eV, respectively, which are similar to the electron temperature.
Note that the slight difference between $T_\mathrm{e}$ and the exponent in the population is caused by a small violation of the local thermal equilibrium in plasma \cite{Fujimoto}.
Based on approximate electron impact excitation and deexcitation rates, we find that this effective temperature has a weak $n_\mathrm{e}$ dependence and approaches $T_\mathrm{e}$ in large $n_\mathrm{e}$ limit.
Even with a smaller electron density, $n_\mathrm{e} = 10^{17}\;\mathrm{m^{-3}}$, this temperature is expected to be $\approx 0.7\ T_\mathrm{e}$ for the $T_\mathrm{e} = 0.34$~eV case.
Details can be found in the Supplementary Material \cite{Note2}.

The histograms in Fig.~\ref{fig:intensity}(a) show the state density $\rho_n(n)$ with given population $n$ (but scaled by $n$ to aid visualization).
The solid blue and red lines are computed according to \eref{eq:derivation3} with $T_\mathrm{e} =$ 0.32 and 0.61 eV, respectively (the same temperatures used in Fig. \ref{fig:level_energy}(b)).
Their agreement is clear.

Figure \ref{fig:intensity}(b) shows the line intensity distribution $\rho_I(I)$ in the visible and infrared wavelength range (scaled by $I$ for visualization).
The solid lines show \eref{eq:final} with $T_\mathrm{e}=$ 0.32 and 0.61 eV for the two cases.
This also agrees with the above discussion, particularly in the first three orders studied by Learner.

\begin{figure*}
    \centering
    \includegraphics[width=17cm]{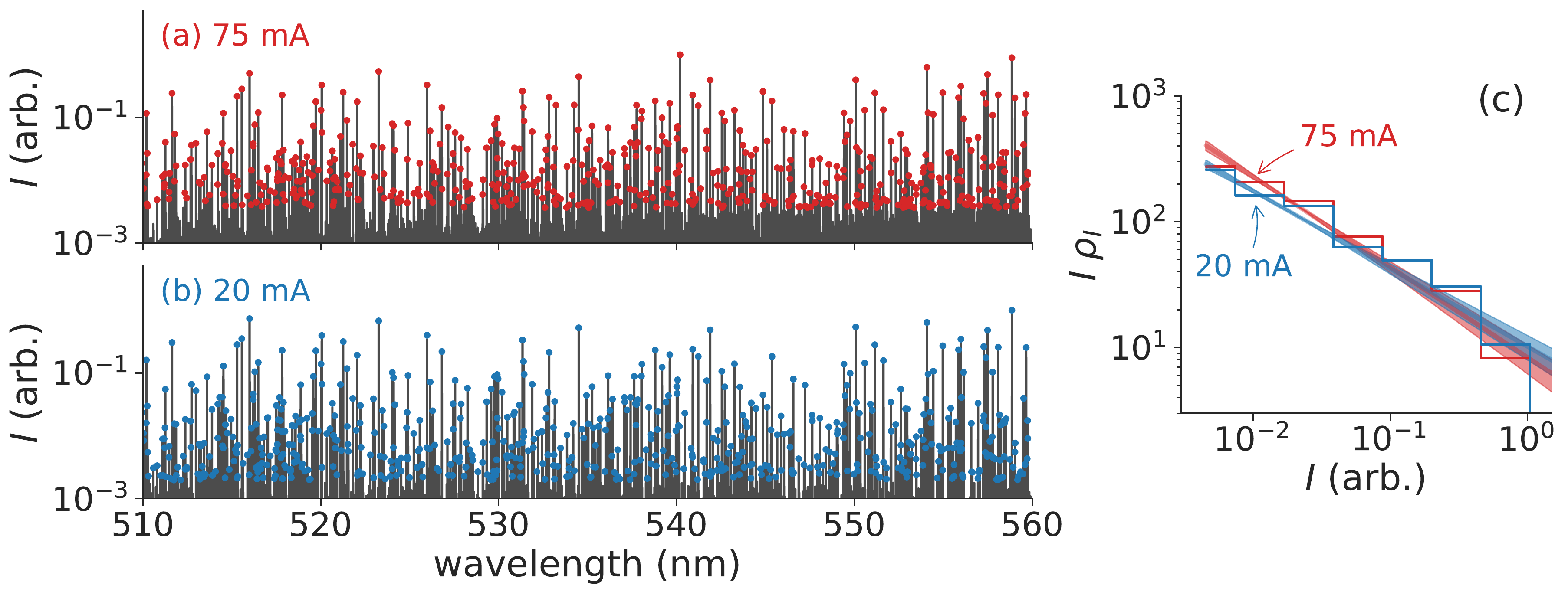}
    \caption{Emission spectra observed from thorium-argon hollow cathode plasmas (a) with 75 mA disharge current \cite{Palmer1980} and (b) with 20 mA discharge current \cite{Kerber2008}.
    Dots indicate the centers and intensities of emission lines detected in each spectrum.
    (c) Intensity distributions computed from the spectra in (a) and (b).
    Straight lines show the optimized power law distribution by maximum likelihood method.
    The exponents for the 75 mA and 20 mA discharges are $1.71 \pm 0.03$ and $1.64 \pm 0.03$, respectively.
    }
    \label{fig:thorium}
\end{figure*}

In \eref{eq:final}, we show that the exponent exclusively depends on $T_\mathrm{e}$ and $\epsilon_0$, but not on other atomic data such as level energies, transition rates, and collision cross sections.
The only value we require, $\epsilon_0$, is known to be accurately calculated with several atomic structure packages \cite{Dzuba2010}.
Therefore, \eref{eq:final} may be useful as a quick diagnostic method for many-electron atom plasmas.

As \eref{eq:final} is scale-free for $I$, the power law dependence is not affected by the system's sensitivity (see the Supplementary Material for details~\cite{Note2}).
Thus, no system calibration is required for estimation of $T_\mathrm{e}$.
We only need to know the dominant (in terms of the number of emission lines) atom in the plasma.

We have applied this approach to the emission spectra measured for thorium plasmas.
Figure \ref{fig:thorium} (a) and (b) show spectra measured from a thorium-argon hollow cathode plasma with 75 mA discharge current~\cite{Palmer1980} and 20 mA discharge current~\cite{Kerber2008}, respectively, with a 1-m Fourier transform spectrometer.
The original data can be downloaded from Kitt Peak National Observatory website~\cite{Kittpeak}.
Only the spectra in 510--560 nm wavelength range are shown and analyzed in this work.
Dots in each panel show the line centers and peaks detected in the two spectra.

Although not all emission lines in these spectra have been identified, we assume that most of the lines are from neutral thorium.
We compute $\rho_I(I)$ from all line intensities in the wavelength range (Fig. \ref{fig:thorium}(c)).
The two histograms generated from the spectra show a power-law distribution.
$\rho_I(I)$ for the higher current discharge shows a steeper slope.

We estimate the exponent of these distributions using the maximum-likelihood method.
The optimized distributions are shown by solid lines (and their 2-$\sigma$ uncertainty by colored bands) in Fig. \ref{fig:thorium} (c).
They fit both histograms.
Estimated values of the exponent are $1.71 \pm 0.03$ and $1.64 \pm 0.03$ for the 75 mA and 20 mA disharges, respectively.
From \eref{eq:final} and the value $\epsilon_0 \approx 0.68$ eV for neutral thorium by Dzuba \emph{et al.} \cite{Dzuba2010}, electron temperatures for these plasmas are estimated as $0.24 \pm 0.01$ eV and $0.21 \pm 0.01 $ eV, respectively.
A higher $T_\mathrm{e}$ value is estimated for the higher current discharge.
Although the positive current dependence of the temperature is not trivial \cite{Cottereau1979}, this dependence qualitatively supports our model.
Because there are no radiative rates reported for neutral thorium, it is difficult to estimate $T_\mathrm{e}$ for this plasma by conventional methods.
To our knowledge, the above procedure is the only one available to estimate $T_\mathrm{e}$ for thorium plasmas.

Although there are significant demands to diagnose plasmas with many-electron atoms,
quantitative comparison with an \textit{ab initio} computer simulation model is not yet accurate enough, because of the unavailability of accurate atomic data.
Our result suggests a possibility of plasma diagnostics that requires only the energy level statistics and the emission intensity statistics. 
Although the validity of the local thermal equilibrium assumption should be investigated further, this may open the door to a new statistical plasma spectroscopy.

In summary, we have presented a simple explanation of Learner's law, where the histogram of the emission line intensities from many-electron atoms follows a power law.
We observed that the exponent is analytically represented with $T_\mathrm{e}$ and $\epsilon_0$.
A similar discussion should also be applicable to other fermionic many-body systems as long as the two assumptions are satisfied.
Although as yet there are no reports about the emission statistics except for many-electron atoms, it is interesting to investigate other systems, such as heavy nuclei.
This is left for future study.

\begin{acknowledgments}
This work was partly supported by JSPS KAKENHI Grant Number 19K14680, the grant of Joint Research by the National Institutes of Natural Sciences (NINS). (NINS program No, 01111905), and partly by the Max-Planck Society for the Advancement of Science. JCB is supported by the Alexander von Humboldt Foundation. We thank Jos\'e Crespo L\'opez-Urrutia, Tomoyuki Obuchi, and Akira Nishio for useful discussions.

\end{acknowledgments}

\bibliography{refs} 

\pagebreak
\widetext
\begin{center}
\textbf{\large A Simple Explanation for the Observed Power Law Distribution of Line Intensity\\
in Complex Many-Electron Atoms
}
\end{center}
\setcounter{equation}{0}
\setcounter{figure}{0}
\setcounter{table}{0}
\makeatletter
\renewcommand{\theequation}{S\arabic{equation}}
\renewcommand{\thefigure}{S\arabic{figure}}

\section{Detailed Derivation of the Intensity Distribution}

In the main text, we used ``$\propto$" in most of the equations and ignored constant factors (e.g., Eqs. 2, 3, and 6) to simplify the discussion.
Furthermore, we made a rather drastic approximation in Eq. (8):   
the constant radiative transition rate.
In this Supplemental Material, we present explicit formulae and explain the approximation in more detail.
As we will see, even if we consider the probability distribution of the transition rate, we arrive at the same result.

Let us redefine explicitly the level density $\rho_E(E)$ and the population $n_i$ at state $i$ having the excited energy $E_i$ as follows,
\begin{equation}
    \label{sup:eq:rhoE}
    \rho_E(E) = \rho_0 \exp\left(\frac{E}{\epsilon_0}\right),
\end{equation}
and
\begin{equation}
    \label{sup:eq:boltzmann}
    n_i = n_0 \overline{g} \exp\left(-\frac{E_i}{kT_\mathrm{e}}\right),
\end{equation}
where $\rho_0$ and $n_0$ are constants.
$\overline{g}$ is the averaged statistical weight for all the state, which we assumed constant over all the levels in the main text.
The distribution of $g_i$ does not affect the result as long as the distribution is uniform and independent of energy.

The number of states having the population $n\sim n+\mathrm{d}n$ can be written as
\begin{align}
    \rho_n(n)\mathrm{d}n &= \rho_E(E)\mathrm{d}E =
    \frac{kT_\mathrm{e}}{n}\rho_E(E)\mathrm{d}n\\
    &= \frac{kT_\mathrm{e}}{n}\rho_0 \exp\left(-\frac{k T_\mathrm{e}}{\epsilon_0} \left(
        \log n - \log n_0 - \log \overline{g} \right)\right)\mathrm{d}n \\
    &= \rho_0 kT_\mathrm{e}(n_0\overline{g})^{kT_\mathrm{e}/\epsilon_0} n^{(-kT_\mathrm{e}/\epsilon_0-1)}\mathrm{d}n
\end{align}

Let us consider the number of emission lines found in the photon energy range of $\omega \sim \omega + \mathrm{d} \omega$ from level $i$ with $0 \le \omega \le E_i$.
Because of the finite wavefunction mixing, the number of emission lines equals to the number of levels within $E_i - (\omega + \mathrm{d} \omega) \sim E_i - \omega$, which is $\rho_E(E - \omega) \mathrm{d}\omega$.
The number of emission lines from the excited states existing within the excited energy $E \sim E + \mathrm{d} E$ is
\begin{align}
    \label{sup:eq:num_lines}
    L(E, \omega) \mathrm{d} \omega \mathrm{d} E
    & = \rho_E(E - \omega)\mathrm{d}\omega \rho_E(E) \mathrm{d} E\\
    & = \rho_0^2\exp\left(\frac{2E - \omega}{\epsilon_0}\right)\mathrm{d} \omega \mathrm{d} E
\end{align}

The intensity of the emission line corresponding to the transition from the state $i$ to state $j$, $I_{ij}$, is written as $I_{ij} = A_{ij} n_i$, where $A_{ij}$ is the radiative transition rate from state $i$ to state $j$.
$A_{ij}$ relates to the line strength $S_{ij}$, $A_{ij} = \gamma\omega_{ij}^3 S_{ij}$, with $\omega_{ij} = E_i - E_j$ and $\gamma = \frac{16\pi^3e^2}{3\nu_0h^4c^3}$, with $e$ the elementary charge, $\nu_0$ vacuum permittivity, $h$ is Planck's constant, and $c$ light speed.

In order to simplify the discussion, for the moment let us assume $A_{ij}$ has a solid relation $A_{ij} = \gamma \omega_{ij}^3 S_0$ for all $i$ and $j$ pairs with a constant $S_0$, and relax this assumption later.
In this case the intensity only depends on the population of state $i$.
The number of lines with given intensity $I_\mathrm{c}$ under this constant $S_0$ assumption is
\begin{align}
    \label{sup:eq:constant}
    \rho_{I\mathrm{c}}(I_\mathrm{c}, \omega)\mathrm{d} I_\mathrm{c} \mathrm{d} \omega
        &= L(E, \omega) \mathrm{d} E \mathrm{d} \omega\\
        &= \rho_0^2\exp\left(-\frac{\omega}{\epsilon_0}\right)
          \exp\left(-2 \frac{kT_\mathrm{e}}{\epsilon_0} \log\left(\frac{I_\mathrm{c}}{\omega^3 \gamma S_0 n_0 \overline{g}}\right)\right) \frac{k T_\mathrm{e}}{I_\mathrm{c}}\mathrm{d} I_\mathrm{c}  \mathrm{d} \omega \\
        & = \rho_0^2\exp\left(-\frac{\omega}{\epsilon_0}\right)
          \left(\omega^3 \gamma S_0 n_0 \overline{g}\right)^{(2kT_\mathrm{e} / \epsilon_0)}
          k T_\mathrm{e}
          I_\mathrm{c}^{(-2kT_\mathrm{e} / \epsilon_0 - 1)} \mathrm{d} I_\mathrm{c} \mathrm{d} \omega,
\end{align}
where $\mathrm{d} I_\mathrm{c}$ is the intensity range that corresponds to the energy range $\mathrm{d} E$, which can be found from \eref{sup:eq:boltzmann} and $I_\mathrm{c} = \omega^3 \gamma S_0 n$. Note here already that the dependence of $I_c$ in the distribution follows the final distribution~(Eq. (2)).

Now we relax the constant line strength assumption and consider the stochastic nature of $S_{ij}$.
Let $\tilde{s}$ be the fluctuation in $S$, $S = \tilde{s} S_0$.
The probability distribution of $\tilde{s}$ may be written as $p_{\tilde{s}}=\frac{1}{\sqrt{2 \pi \tilde{s}}}\exp(-\tilde{s}/2)$, which is a Porter-Thomas distribution with mean 1 \cite{Porter1956, Grimes1983, Bisson1991, Flambaum1998}.
As we will see later, the actual shape of $p_{\tilde{s}}$ is not important as long as it decays exponentially in the large $\tilde{s}$ limit.
More importantly, we here assume that $p_{\tilde{s}}$ is independent and identical for all the transitions.
With this assumption, the intensity $I$ becomes a product of two independent random variables, $I = \tilde{s} I_\mathrm{c}$.

Let us consider only the probability distribution of $I_\mathrm{c}$, rather than the number of lines.
Since the distribution of $I_\mathrm{c}$ follows the power law, the probability distribution of $I_\mathrm{c}$ is written as a Pareto distribution,
\begin{equation}
    p_{I_\mathrm{c}}(I_\mathrm{c}) = \alpha I_\mathrm{min}^\alpha I_\mathrm{c}^{-\alpha - 1}\;\; (\mathrm{s.t.}\; I_\mathrm{c} \ge I_\mathrm{min})
\end{equation}
where $\alpha = 2 k T_\mathrm{e} / \epsilon_0$ and $I_\mathrm{min}$ is the minimum value of the intensity.
We will take a limit of $I_\mathrm{min} \to 0$ later but in order to make the distribution integrable, we now assume this value is sufficiently small.
The probability distribution of the intensity is written as
\begin{align}
    \label{sup:eq:final}
    p(I | \omega) &= \int_0^\infty p_{\tilde{s}}(\tilde{s}) p_{I_\mathrm{c}}(I / \tilde{s}) \frac{1}{\tilde{s}} \mathrm{d}\tilde{s} \\
                  &= \alpha I_\mathrm{min}^\alpha I^{(-\alpha-1)}
                  \int_0^{I/I_\mathrm{min}} \tilde{s}^\alpha p_{\tilde{s}}(\tilde{s}) \mathrm{d}\tilde{s}.
\end{align}
If $p_{\tilde{s}}(\tilde{s})$ decays faster than ${\tilde{s}}^{-\alpha-1}$ in large $\tilde{s}$ limit (such as the exponential decay) and $I \gg I_\mathrm{min}$, the integral in the right hand side converges and does not depend on $I$.
With sufficiently small $I_\mathrm{min}$ (which corresponds to sufficiently large $E$ for the upper state), $p(I | \omega)$ becomes the power law distribution except for the very small $I$ region.
This distribution actually has the same shape as \eref{sup:eq:constant} where we assume that $S$ is constant.

Let us consider the sensitivity of the observation system, $\xi(\omega)$.
The density distribution of the observed intensity $I' = \xi(\omega)I$ is
\begin{align}
  \label{sup:eq:observed}
  \rho_{I'}(I')\mathrm{d}I' = \rho_0^2\exp\left(-\frac{\omega}{\epsilon_0}\right)
  \left(\omega^3 \gamma S_0 n_0 \overline{g} \xi(\omega)^{-1}\right)^{(2kT_\mathrm{e} / \epsilon_0)}
  k T_\mathrm{e}
  I'^{(-2kT_\mathrm{e} / \epsilon_0 - 1)} \mathrm{d} I'.
\end{align}
The system's sensitivity only affects the scale of the distribution, but not the exponent of the power law.

A histogram may be computed from an emission spectrum observed in finite wavelength range, $\Omega$.
It corresponds to the integration of \eref{sup:eq:observed} over the observation wavelength range,
\begin{align}
 \int_\Omega\mathrm{d}\omega \rho_{I'}(I')\mathrm{d}\omega\mathrm{d}I'
 \propto I'^{-2kT_\mathrm{e} / \epsilon_0 - 1}\mathrm{d}I'
\end{align}
which results in the same power law.
Therefore, no system calibration is necessary to compute the intensity histogram.
This ergodic property essentially comes from the fact that the power law is scale invariant.

\section{Bias in the $T_\mathrm{e}$ estimation}

The $T_\mathrm{e}$ estimation method newly proposed in this work is based on the local thermal equilibrium assumption.
As a slight temperature difference can be found between the simulated and estimated $T_\mathrm{e}$ values in the main text, this assumption is not always satisfied perfectly.
Here, we present a simple analytical model to predict the estimation bias.

We consider plasmas with weak radiation field, where the photo excitation of atoms is negligible.
We further neglect ionization and recombination for the sake of the simplicity, and only consider electron collision excitation and deexcitation, as well as radiative deexcitation.
Let us assume that the excited level population of an atom is approximated by Boltzmann's distribution with an effective temperature $T_\mathrm{eff}$,
\begin{equation}
    \label{sup:eq:boltzmann_eff}
    n_i = n_0 \overline{g} \exp\left(-\frac{E_i}{kT_\mathrm{eff}}\right),
\end{equation}
where $T_\mathrm{eff}$ may be different from $T_\mathrm{e}$.
We seek the value of $T_\mathrm{eff}$ where the population balance is established under electron-collisional excitations, deexcitations and radiative deexcitations.

Let $r_{ij}(T_\mathrm{e})$ be the collisional excitation rate from $i$ to $j$ states.
Based on the detailed balance principle, the inverse process, i.e., the collisional deexcitation rate from $j$ to $i$ state can be written as follows,
\begin{equation}
    \label{sup:sub:detailed_balance}
    r_{ij}(T_\mathrm{e}) = r_{ji}(T_\mathrm{e}) \exp\left(-\frac{\omega_{ij}}{kT_\mathrm{e}}\right),
\end{equation}
where $\omega_{ij}$ is the energy difference between $i$ and $j$ states.
Here, we neglect the statistical weight difference in states $i$ and $j$ as in the previous section.
The net population influx from $j$ to $i$ states is written as
\begin{align}
    \label{sup:eq:net_collisional_influx}
    \gamma_{ji}^\mathrm{e} 
    &= - n_i r_{ij}(T_\mathrm{e}) + n_j r_{ji}(T_\mathrm{e}) \\
    \label{sup:eq:net_collisional_influx2}
    &= 
    n_0 \overline{g}\exp\left(-\frac{E_i}{kT_\mathrm{eff}}\right)
    r_{ij}(T_\mathrm{e})
    \left\{
        \exp\left(-\left(
            \frac{1}{kT_\mathrm{eff}} - \frac{1}{kT_\mathrm{e}}\right)\omega_{ij}\right)
        - 1
    \right\}.
\end{align}

The total population flux into state $i$ is $\Gamma_i^\mathrm{e} = \sum_j \gamma_{ji}^\mathrm{e}$.
We approximate it by continous integration taking the level density $\rho_n$ into account,
\begin{equation}
    \label{sup:eq:collisional_influx}
    \Gamma_i^\mathrm{e} = 
    \int_0^\infty \gamma_{i}^\mathrm{e}(E_i + \omega) \rho_n(E_i + \omega) \mathrm{d}\omega + 
    \int_0^{E_i} \gamma_{i}^\mathrm{e}(E_i - \omega) \rho_n(E_i - \omega) \mathrm{d}\omega,
\end{equation}
where $\gamma_{i}^\mathrm{e}(E_j) = \gamma_{ji}^\mathrm{e}$ if $E_j > E_i$ otherwise $\gamma_{i}^\mathrm{e}(E_j) = -\gamma_{ij}^\mathrm{e}$.

We use an approximate form of $r_{ij}$,
\begin{equation}
    \label{sup:eq:approx_rate}
    r_{ij}(T_\mathrm{e}) \approx 
    \frac{\beta}{\sqrt{kT_\mathrm{e}}} S_{ij} \exp\left(-\frac{\omega_{ij}}{kT_\mathrm{e}}\right) n_\mathrm{e},
\end{equation}
with $\beta = \sqrt{\frac{2^9}{3}}\pi^{3/2}m_{\mathrm{e}}^{-1/2}a_0^2 E_{\mathrm{H}}^2 \overline{\xi}
\frac{8\pi^2 m_{\mathrm{e}}}{3h^2}$,
where $m_\mathrm{e}$ the electron mass, $a_0$ Bohr radius, $E_\mathrm{H}$ Rydberg unit of energy, and $\overline{\xi}\approx0.6$ the integrated gaunt factor~\cite{Mewe1972}.
Here, we substituted the relation of the oscillator strength $f_{ij} = \frac{16\pi^{2}}{3e^2 m_\mathrm{e} h^2 c^2 \overline{g}} \omega_{ij} S_{ij}$ to the original formula found in Ref.~\cite{Mewe1972}.
Substituting Eqs. (\ref{sup:eq:approx_rate}) and (\ref{sup:eq:net_collisional_influx2}) into \eref{sup:eq:collisional_influx}, $\Gamma_{i}^\mathrm{e}$ can be written as
\begin{equation}
    \label{sup:eq:collisional_influx2}
    \Gamma_i^\mathrm{e} = n_\mathrm{e} n_0 \overline{g} \beta S_0 \rho_0 \sqrt{kT_\mathrm{e}}
    \exp\left(-\left(
        \frac{1}{k T_\mathrm{eff}} - \frac{1}{\epsilon_0}
    \right)E_i\right)
    \left\{
        \frac{\epsilon_0}{k T_\mathrm{e}}\frac{1}{x}\left(1 - \exp\left(-x \frac{E_i}{\epsilon_0}\right)\right) - 
        \frac{\epsilon_0}{k T_\mathrm{e}}
        - \frac{1}{1 - \frac{kT_\mathrm{e}}{\epsilon_0}}
        + \frac{1}{\frac{kT_\mathrm{e}}{kT_\mathrm{eff}} - \frac{k T_\mathrm{e}}{\epsilon_0}}
    \right\},
\end{equation}
where 
$x = 1 + \frac{\epsilon_0}{kT_\mathrm{e}} - \frac{\epsilon_0}{kT_\mathrm{eff}}$.
Note that we neglected the variation of $S_{ij}$ and assume it as constant $S_0$.
$E_i \gg \epsilon_0$ is also assumed in the above derivation.

The population influx by radiative transition into state $i$ can be written as 
$\Gamma_i^\mathrm{rad} = \sum_{j > i} A_{ji} n_j - \sum_{j < i} A_{ij} n_i$, where $j > i$ ($j < i$) indicates the summation over the state $j$ having with $E_j > E_i$ ($E_j < E_i$).
We approximate it by a continuous integration as follows:
\begin{align}
    \label{sup:eq:radiative_influx}
    \Gamma_i^\mathrm{rad} 
    &= 
    \int_0^\infty n_0 \overline{g} \exp\left(-\frac{E_i + \omega}{k T_\mathrm{eff}}\right)
    \rho(E_i + \omega) \gamma \omega^3 S_0 \mathrm{d}\omega 
    - 
    \int_0^{E_i} n_0 \overline{g} \exp\left(-\frac{E_i}{k T_\mathrm{eff}}\right)
    \rho(E_i - \omega) \gamma \omega^3 S_0 \mathrm{d}\omega \\
    &= 6 n_0 \overline{g} \gamma S_0 \rho_0 \exp\left(-\left(
        \frac{1}{kT_\mathrm{eff}} - \frac{1}{\epsilon_0}
    \right)E_i \right)
    \left\{
        \left(\frac{1}{\frac{1}{kT_\mathrm{eff}} - \frac{1}{\epsilon_0}}\right)^4
        - \epsilon_0^4
    \right\}.
\end{align}

From the steady state condition $\Gamma_i^\mathrm{e} + \Gamma_i^\mathrm{rad} = 0$, $T_\mathrm{eff}$ should satisfy the relation
\begin{align}
    \label{sup:eq:final_Teff}
    \beta n_\mathrm{e} (kT_\mathrm{e})^{-7/2}\left\{
        \frac{\epsilon_0}{k T_\mathrm{e}}\frac{1}{x}\left(1 - \exp\left(-x \frac{E_i}{\epsilon_0}\right)\right) - 
        \frac{\epsilon_0}{k T_\mathrm{e}}
        - \frac{1}{1 - \frac{kT_\mathrm{e}}{\epsilon_0}}
        + \frac{1}{\frac{kT_\mathrm{e}}{kT_\mathrm{eff}} - \frac{k T_\mathrm{e}}{\epsilon_0}}
    \right\}
    + 
    6 \gamma \left\{
        \left(\frac{1}{\frac{k T_\mathrm{e}}{k T_\mathrm{eff}} - \frac{k T_\mathrm{e}}{\epsilon_0}}\right)^4
        - \frac{\epsilon_0}{kT_\mathrm{e}}^4
    \right\}
    = 0.
\end{align}
Although its solution is not analytically tractable, it can be easily solved numerically.
We note that the dependence on $E_i$ is not significant if $E_i$ is a few times larger than $\epsilon_0$. 
We assume $E_i \approx 4 \epsilon_0$ in the following numerical evaluation, which corresponds to the first ionization energy of neutral iron $\chi_\mathrm{Fe} = 7.90$ eV with $\epsilon_0 =$ 1.97 eV.
It should be also noted that $T_\mathrm{eff} \lesssim \epsilon_0 / 2$ is required to keep the approximations valid that are used for deriving \eref{sup:eq:final_Teff}.

\begin{figure}
    \centering
    \includegraphics[width=13cm]{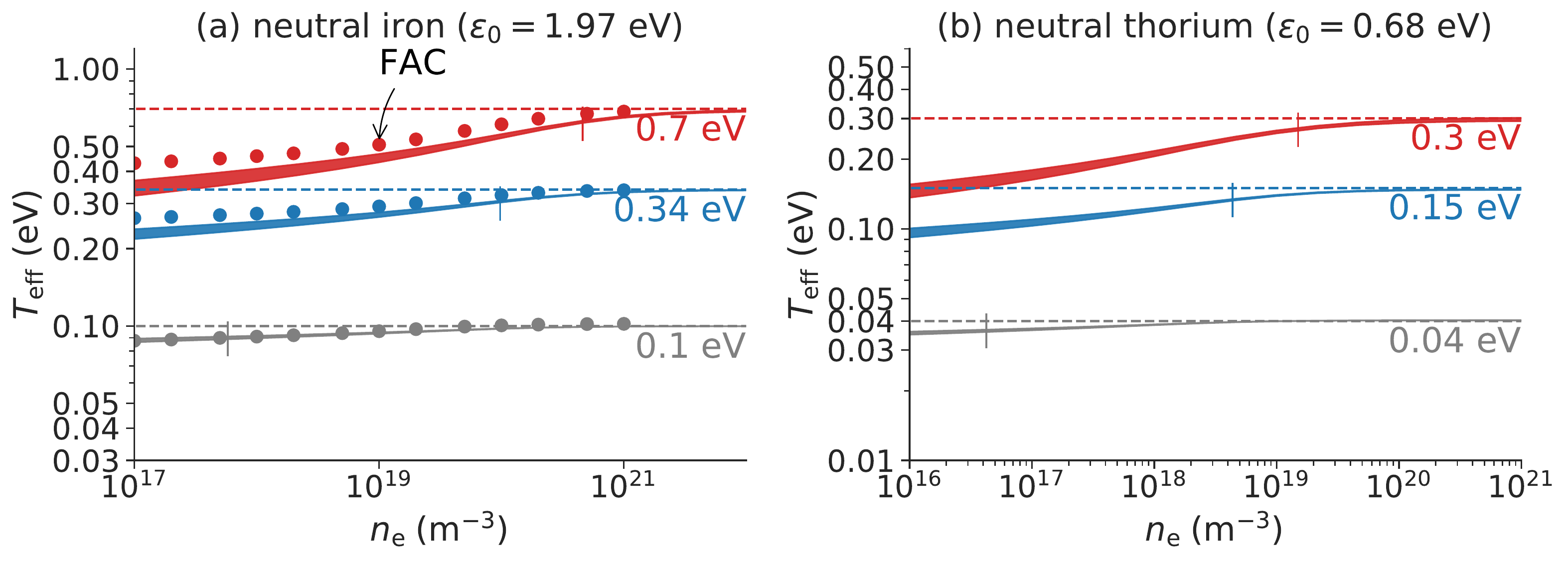}
    \caption{$n_\mathrm{e}$-dependence of $T_\mathrm{eff}$ computed for several $T_\mathrm{e}$ values.
    (a) Results for neutral iron, $\epsilon_0 = 1.97$ eV. (b) Results for neutral thorium, $\epsilon_0 = 0.68$ eV.
    The dependence on $E_i$ is shown by the thickness of the curve, where we changed $E_i$ from $3.5\,\epsilon_0$ to $5\,\epsilon_0$.
    The vertical bars indicate the threshold density where $T_\mathrm{eff} = 0.9\, T_\mathrm{e}$.
    Markers in (a) are the results with FAC.
    \label{sup:fig:Teff}
    }
\end{figure}

The $n_\mathrm{e}$ dependence of $T_\mathrm{eff}$ with $\epsilon_0 = 1.97$ eV, which is the same value used in the main text for neutral iron, is shown in Fig. \ref{sup:fig:Teff} (a).
$T_\mathrm{eff}$ approaches to $T_\mathrm{e}$ in the large
 $n_\mathrm{e}$ limit, i.e., when the relative importance of the radiative decay becomes negligible.
$T_\mathrm{eff}$ decreases in a lower $n_\mathrm{e}$ plasma.
The vertical bars in the figure show threshold densities, at which $T_\mathrm{eff} = 0.9\, T_\mathrm{e}$.
Lower $T_\mathrm{e}$ values give lower threshold densities, however even below the threshold, the $n_\mathrm{e}$ dependence of $T_\mathrm{eff}$ is weak.

We also carry out \textit{ab initio} simulations with FAC under several $n_\mathrm{e}$ and $T_\mathrm{e}$ conditions.
$T_\mathrm{eff}$ is estimated from the simulated excited state population.
The estimated $T_\mathrm{eff}$ values are shown by markers in Fig. \ref{sup:fig:Teff} (a).
Note that FAC utilizes the distorted wave approximation for simulating the electron impact excitation rate, 
which takes the mixing of wavefunctions into account, i.e., this approximation is more robust than \eref{sup:eq:approx_rate}.

A qualitative agreement between the analytical model and FAC is clear, especially in the threshold density.
This indicates the validity of the above approximations.
Although there is still a slight discrepancy in low $n_\mathrm{e}$ region, the agreement is surprising because only $\epsilon_0$ is used as an atomic data in the analytical model.
We may even be able to correct the estimation bias on $T_\mathrm{e}$ based on this model if we have a rough value of $n_\mathrm{e}$ in the plasma.

In Fig. \ref{sup:fig:Teff} (b), we show $T_\mathrm{eff}$ values for neutral thorium with $\epsilon_0 = 0.68$ eV.
For a realistic electron density in the plasma ($n_\mathrm{e} = 10^{18}$ -- $10^{20} \mathrm{\;m^{-3}}$) and for $T_\mathrm{eff} \approx$ 0.2~eV as estimated in the main text, $T_\mathrm{eff}$ is within 30\% of $T_\mathrm{e}$.

This small $n_\mathrm{e}$ dependence indicates the sensitivity for the temperature measurement.
Since the emissivity is roughly proportional to $n_\mathrm{e}$, if $n_\mathrm{e}$ varies over the measurement region and time but $T_\mathrm{e}$ does not, the exponent will correctly represent the temperature.
This is a unique feature of this plasma diagnostic technique with many-electron atom emission, while many of the existing methods must assume a single density and temperature over the measurement range and duration.

\section{Criterion of local thermal equilibrium for many-electron atoms}

For hydrogen-like atomic ions, Griem has proposed a validity criterion of local thermal equilibrium \cite{Griem},
\begin{equation}
    \label{sup:eq:griem}
    n_\mathrm{e} \gtrsim 7 \times 10^{24}\, p^{-17/2} z^7
    \left(\frac{kT_\mathrm{e}}{z^2\chi_\mathrm{H}}\right)^{1/2} \;\mathrm{[m^{-3}]}
\end{equation}
where $z$ the nuclear charge, $p$ the principal quantum number, and $\chi_\mathrm{H} = 13.6$ eV the ionization energy of neutral hydrogen.
In this section, we derive a similar criterion for many-electron atoms from \eref{sup:eq:final_Teff}.

We define the boundary density where $T_\mathrm{eff} \gtrsim 0.9 T_\mathrm{e}$.
With the assumption $0.1 \ll kT_\mathrm{eff} / \epsilon_0 < 1 / 2$, \eref{sup:eq:final_Teff} may be reduced to
\begin{equation}
    \label{sup:eq:criterion}
    n_\mathrm{e} \gtrsim 6 \gamma \beta^{-1}
    (k T_\mathrm{e})^{7/2} \left(\frac{T_\mathrm{e}}{T_\mathrm{e} - T_\mathrm{eff}}\right)
    \left(\frac{\epsilon_0}{kT_\mathrm{e}}\right)^2
    = 2\times 10^{24} 
    \left(\frac{\epsilon_0}{\chi_\mathrm{H}}\right)^{7/2}
    \left(\frac{k T_\mathrm{e}}{\epsilon_0}\right)^{3/2}
    \;\; \mathrm{[m^{-3}]}.
\end{equation}
For the $T_\mathrm{e} = 0.7$ eV case, this boundary density is $2 \times 10^{21} \mathrm{\;m^{-3}}$, which is similar to the numerically derived one ($5 \times 10^{20} \mathrm{\;m^{-3}}$, shown by vertical bars in Fig. \ref{sup:fig:Teff}).

This criterion has a different $T_\mathrm{e}$ dependence to the Griem criterion \eref{sup:eq:griem}.
However we can make a connection with the Griem criterion by noting that the emitter charge state distribution will itself change with temperature. In this case $\epsilon_0$ changes with $T_\mathrm{e}$, but $\frac{\epsilon_0}{kT_\mathrm{e}}$ and $\frac{z^2 \chi_\mathrm{H}}{kT_\mathrm{e}}$ may be approximately constant. Here $z$ is promoted to an ``effective charge'' such that $z^2\chi_\mathrm{H}$ is the ionization energy of the emitter.
From the case with neutral iron and $T_\mathrm{e} = 0.7$ eV, $\frac{\epsilon_0}{kT_\mathrm{e}} \approx 3$ and $\frac{z^2 \chi_\mathrm{H}}{kT_\mathrm{e}} \approx 10$.
Since the chaotic behavior in many-electron atoms starts around $E \gtrsim \epsilon_0$,
the corresponding effective principal quantum number for the chaotic states in neutral iron may be $p \approx \sqrt{\chi_\mathrm{H} / (\chi_\mathrm{Fe} - \epsilon_0)} = 1.5$.
With this effective principal quantum number, Griem's boundary density for many-electron atoms becomes
\begin{equation}
\label{sup:eq:griem_eps}
    n_\mathrm{e} \gtrsim 2 \times 10^{22} T_\mathrm{e}^{7/2}
    \;\; \mathrm{[m^{-3} eV^{7/2}]}
\end{equation}
and our criterion reads
\begin{equation}
    \label{sup:eq:criterion_simple}
    n_\mathrm{e} \gtrsim 2 \times 10^{21} T_\mathrm{e}^{7/2}
    \;\; \mathrm{[m^{-3} eV^{7/2}]},
\end{equation}
with $T_\mathrm{e}$ in eV.
Both the boundary densities scale as $T_\mathrm{e}^{7/2}$ consistently, but they should be understood as providing only a rough order-of-magnitude estimate.

By contrast, the consistency with the criterion proposed by McWhirter \cite{McWhirter}, which is based on the maximum level spacing, is not obvious, because our result does not depend on the absolute level spacing $\approx 1/\rho_0$, i.e., $\rho_0$ is canceled out in \eref{sup:eq:final_Teff}. 
More precise and quantitative discussion of the validity condition for local thermal equilibrium in many-electron atoms is left to future study.


Let us consider the validity of local thermal equilibrium in some typical many-electron atom plasmas.
In laser produced plasmas for ultraviolet light sources, the temperature is typically $\approx 10^2 \mathrm{\;eV}$ while the density varies in the range $10^{25} - 10^{27} \mathrm{\; m^{-3}}$, depending on the incident laser wavelength \cite{Suzuki2017,Osullivan2015}.
This parameter range is around the boundary \eref{sup:eq:criterion_simple}.
Although many parameters are unknown for the laser produced plasmas, including the actual value of $\epsilon_0$ for the dominant emitter ion, the density dependence on the spectral shape (showing a broader spectrum from more dense plasmas) has been observed \cite{Suzuki2017}.

The divertor plasmas of tokamak nuclear fusion reactors is also close to the boundary ($T_\mathrm{e} \approx 10^0 \mathrm{\; eV}, n_\mathrm{e} \approx 10^{20} \mathrm{\;m^{-3}}$) \cite{Kawashima2009, Kukushkin2011}.
On the other hand, parameters of tokamak core plasmas stay typically very far from the threshold, $n_\mathrm{e} \approx 10^{20} \mathrm{\;m^{-3}}$ and $T_\mathrm{e} \approx 10^4 \mathrm{\;eV}$; local thermal equilibrium is not expected for the core plasma.
\end{document}